\documentclass[a4paper, 10pt, conference]{IEEEtran}

\usepackage{url}
\usepackage{graphicx}
\usepackage{amsfonts}
\usepackage{flushend} 
\usepackage{caption}
\usepackage{subcaption}
\usepackage{color}
\usepackage{stfloats}
\usepackage{color,soul}
\usepackage{amsmath}
\usepackage{multirow}
\usepackage{hhline}
\usepackage{gensymb}
\usepackage{xcolor}

\IEEEoverridecommandlockouts                              

\title{\LARGE \bf
Micro-Operator driven Local 5G Network Architecture for Industrial Internet
}

\author{\IEEEauthorblockN{Yushan Siriwardhana\IEEEauthorrefmark{1},
Pawani Porambage\IEEEauthorrefmark{1},
Madhusanka Liyanage\IEEEauthorrefmark{1},
Jaspreet Singh Walia\IEEEauthorrefmark{2},\\
Marja Matinmikko-Blue\IEEEauthorrefmark{1}, Mika Ylianttila\IEEEauthorrefmark{1},}
\IEEEauthorrefmark{1}Centre for Wireless Communications, University of Oulu, Finland. email: \{firstname.lastname\}@oulu.fi\\
\IEEEauthorblockN{\IEEEauthorrefmark{2}Aalto University, Finland. email: jaspreet.walia@aalto.fi }
}

\begin{document}

\maketitle
\thispagestyle{empty}
\pagestyle{empty}

\begin{abstract}
In addition to the high degree of flexibility and customization required by different vertical sectors, 5G calls for a network architecture that ensures ultra-responsive and ultra-reliable communication links. The novel concept called micro-operator~(uO) enables a versatile set of stakeholders to operate local 5G networks within their premises with a guaranteed quality and reliability to complement mobile network operators' (MNOs) offerings.  In this paper, we propose a descriptive architecture for emerging 5G uOs which provides user specific and location specific services in a spatially confined environment. The architecture is discussed in terms of network functions and the operational units which entail the core and radio access networks in a smart factory environment which supports industry 4.0 standards. Moreover, in order to realize the conceptual design, we provide simulation results for the latency measurements of the proposed uO architecture with respect to an augmented reality use case in industrial internet. Thereby we discuss the benefits of having uO driven local 5G networks for specialized user requirements, rather than continuing with the conventional approach where only MNOs can deploy cellular networks. 

\end{abstract}

\begin{IEEEkeywords}
Micro-operators, Industrial Internet, Industry 4.0, 5G, Augmented Reality, Architecture 
 
\end{IEEEkeywords}

\section{Introduction}

The landscape of mobile communication service requirements is rapidly changing with the proliferation of digitization technologies. Consequently, in the future, more emphasis needs to be placed on location specific services in different vertical sectors. Hospitals, shopping malls, smart cities, industries and universities are identified as some of the common locations, which are heavily benefited by these location specific services. Location specific requirements stipulate high demands on reliability, high data rates, low latency, privacy and security. The key focus of the future 5G wireless systems is to serve such case specific requirements along with the provisioning of the traditional mobile broadband services~\cite{euroadmap}. These case specific and localized requirements are expanding beyond the current capabilities of the traditional MNOs whose services are often designed to serve masses. To cater the location specific future communication requirements, the need for establishing local 5G networks is evident. In speeding up local service delivery with 5G networks, the present mobile communication market needs to be opened for local 5G networks deployed by different stakeholders such as recently proposed in the micro operator (uO) concept~\cite{matinmikko2017micro}.

Unlike the traditional MNOs with wide area coverage, uOs are local operators who intend to offer case specific and location specific services through locally deployed 5G networks ~\cite{matinmikko2017micro1}. Therefore, system architecture for a uO should be carefully designed in such a way that it enables the efficient and reliable local service deliveries. Since many services have very stringent requirements that can only be provided through cellular network technology, uO itself is a 5G service provider and uO system architecture must contain the network functions defined by 3rd Generation Partnership Project~(3GPP). Since the uOs are specialized to provide tailored services, the system architecture and its specific deployment may also depend on the use case.

Besides the novelty of uO concept, the uO system architecture is still not defined in a comprehensive manner. With this regard, we propose a descriptive architecture for emerging 5G uO, which provides user specific and location specific services in a spatially confined environment. The architecture is discussed for a smart factory environment which supports industry 4.0 standards, having typical use case called Augmented Reality (AR)~\cite{automation}. The architecture comprises 5G network functions and the operational units which entail the core and the access networks to cater the communication of AR use case. In order to realize the conceptual design and present the simulation results, we compare two network deployment models for a factory, one being served by a local uO or and the other being served by a traditional MNO. Based on the simulation results, we discuss the benefits of having uOs in 5G for specialized user requirements, rather than continuing with the traditional MNO driven approach.

The remainder of the paper is organized as follows: Section~\ref{sec:RelatedWork} describes the related work on uOs, generic 5G architecture and industry 4.0. Section~\ref{sec:usecaseindustry} describes the use case for which the uO architecture is defined and Section~\ref{sec:experimentalsetup} presents the experimental setup and its key parameters. Section~\ref{sec:discussion} compares and discusses the simulation results for a typical MNO setting and the proposed uO architecture. Finally, Section~\ref{sec:conclusions} concludes the paper with the future research directions.


\section{Background and Related Work}
\label{sec:RelatedWork}


Expected key characteristics of the future 5G wireless systems are identified as extremely high data rates, ultra reliability and low latency, and massive communication between devices~\cite{shafi20175g}. Moreover, three specific areas of 5G services are diversified as enhanced mobile broadband (eMBB), ultra reliability and low latency communication (URLLC) and massive machine type communication (mMTC)~\cite{itudoc}. Based on the communication needs of different verticals, future 5G operators must possess the capability of providing case specific services in addition to the present generic communications services.

Local 5G networks are gaining increasing attention in regulation, research and industry. The concept of uO was proposed to expand the mobile communication market by allowing new stakeholders to deploy local 5G networks to complement the conventional MNOs. The uOs are expected to provide tailored 5G services and fulfill case specific and versatile local wireless communication needs with extremely low latency~\cite{matinmikko2017micro}. 5G uO can operate a closed network to serve its own customers, an open network to offer its services to other MNO's customers, or a mix of both. Key regulatory elements and the techno economic aspects related to the uOs are discussed in~\cite{matinmikko2017micro1}. Business model options for local 5G uOs and the different network deployment options are discussed in~\cite{ahokangas2016future}.

\begin{figure}[ht]
\centering
    \includegraphics[width=0.45\textwidth]{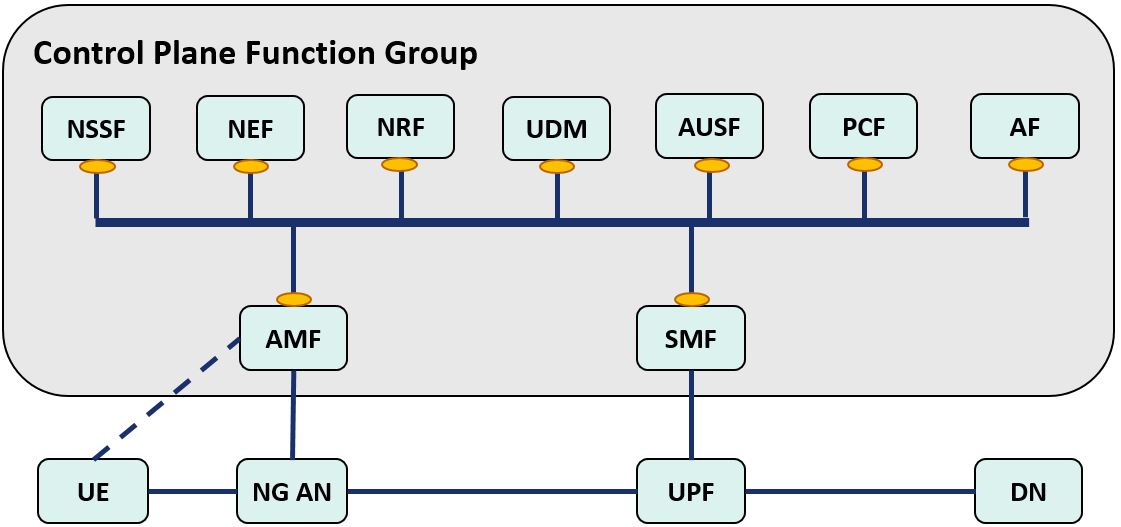}
  \caption{Service based representation of 5G system architecture~\cite{architecture}}
  \label{fig_service}
\end{figure}
\begin{figure}[ht]
\centering
    \includegraphics[width=0.45\textwidth]{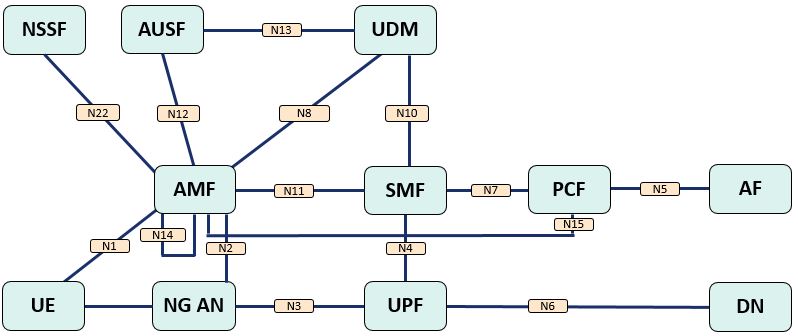}
  \caption{Reference point representation of 5G system architecture~\cite{architecture}}
  \label{fig_point}
\end{figure}

The network architecture of 5G uO should also comprise the network functions of generic 5G architecture~\cite{matinmikko2017micro}. 3GPP has already released the specifications for 5G system architecture~\cite{architecture}. Instead of the network elements defined in Evolved Packet Core (EPC) in 4G systems, Software Defined Networking~(SDN) and Network Function Virtualization~(NFV) are involved in creating Network Functions~(NF) in 5G systems architecture. Network functions can be implemented on a dedicated hardware or as a software instance on a dedicated hardware or as a virtualized function instantiated on an appropriate platform such as a cloud. The concept of network functions has led operators to add flexibility over the functionality of the underlying physical infrastructure of the 5G network. 3GPP specifications represent the 5G architecture in two ways:
\begin{itemize}
    \item \textbf{Service based representation:} Shows how NFs within the control plane enable other authorized NFs to access their services, as in Figure \ref{fig_service}. 
    \item \textbf{Reference point representation:} , Shows the point-to-point interaction existing between two NFs, as depicted in Figure \ref{fig_point}.
\end{itemize}

Different locations belong to different vertical sectors (i.e., hospitals, campuses, factories, etc.) have their own communication requirements, hence the network architecture also depends on the communication requirement. In this paper, we consider Industry 4.0 or smart factory environment to develop the uO architecture. Industry 4.0 refers to the advancement of the present industries into the next generation~\cite{wollschlaeger2017future}. It aims to interconnect the devices inside the factories, make them smart by adding more intelligence into the device and ultimately resulting in improved adaptability, resource efficiency, and the supply and demand process between factories~\cite{varghese2014wireless}. Machine-to-Machine (M2M) communication plays a critical role in Industry 4.0 which is also a key focus in 5G systems. Wireless Sensor Networks (WSN) in current industries are moving towards industrial wireless networks because of the low latency, high mobility and high capacity requirements in the future industries~\cite{li2017review}. A study report has been released by 3GPP focusing on typical use cases in Industry 4.0 such as motion control, mobile robots, augmented reality, massive wireless sensor networks~\cite{automation}.


\section{Use Case and Proposed Architecture}
\label{sec:usecaseindustry}

\begin{figure}[ht]
\centering
    \includegraphics[width=0.47\textwidth]{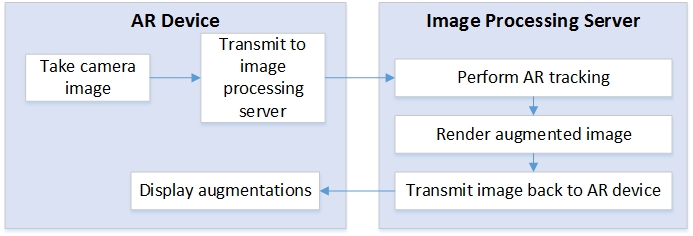}
  \caption{AR system model with offloaded processing~\cite{automation}}
  \label{fig_ar}
\end{figure}

Augmented Reality (AR) can be considered as an application which will be heavily used in future industrial environments~\cite{automation}. In our study, we consider AR use case in which the factory workers are supported by AR devices. They identify production flaws, obtain step by step guidance to carry out pre-defined tasks, obtain support from the supervisors via those devices. In this context, AR devices should be highly energy efficient and lightweight. This requires AR devices to carry out minimal processing and more intensive tasks to be carried out by a separate image processing server located inside the factory. Typical communications of an Industry 4.0 AR network should be as depicted in Figure \ref{fig_ar}.

AR device takes the images and transmits them to the image processing server. Server then does the processing of the images and sends the augmentations back to AR device to display. 5G system supporting this communication should be able to provide an end-to-end latency of less than 10 ms for one way communication with a 99.9\% success of frame delivery~\cite{automation}.

A local 5G network deployed by a uO covering the factory could be used to address the needs of AR use case. Because uO operates a local 5G network, it should comprise architectural components inherited from generic 5G systems architecture. The uO concept provides flexibility over the selection of architectural components and the location where the core network is hosted. For a low latency requirement, the desirable implementation is to have the uO core network within the factory premises itself, but not mandatory.

In our study, we define the architectural components needed in the core network to cater AR use case. Generally, AR use case requires the 5G system facilitate the following three steps of communications.

\begin{itemize}
    \item Registering the AR devices into the network
    \item Establishing data session between the AR device and image processing server 
    \item Data transfer between AR device and image processing server
\end{itemize}

\begin{figure}[ht]
\centering
    \includegraphics[width=0.47\textwidth]{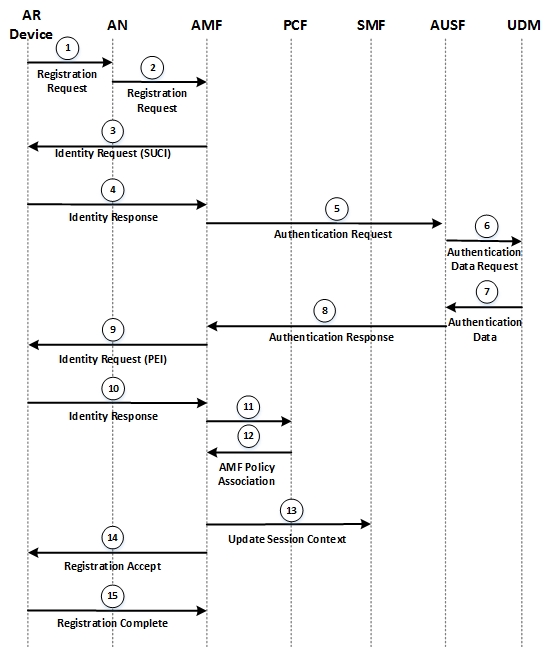}
  \caption{Message sequence chart for AR device registration procedure }
  \label{fig_regis}
\end{figure}

Architectural components needed for completing above steps can be identified based on the message transfer between each element in 5G system including AR device, Next Generation NodeB~(gNB) and core network functions. We define the registration procedure for AR device based on 3GPP specifications~\cite{procedures}. Figure \ref{fig_regis} illustrates the message sequence between the entities in the architecture.

AR device initiates the registration process by sending registration request to gNB. gNB forwards the request to  Access and Mobility Management Function (AMF). After that, AMF and AR device exchange the identity request and response messages. In the next step, AMF contacts Authentication Server Function~(AUSF) for the device authentication. AUSF facilities the authentication after contacting the Unified Data Management~(UDM) and retrieving the authentication data. Once the authentication data is received from UDM, AUSF sends the authentication response to AUSF. Identity request/response messages are transmitted between AMF and the AR device again. After the identity verification, AMF then works with Policy Control function~(PCF) for the policy association for the AR device. Once the policy association is successful, AMF sends an update to Session Management Function~(SMF) informing the session context. AMF also sends the registration accept message to the AR device and the device then sends registration complete message to AMF concluding the registration process.

\begin{figure}[ht]
\centering
    \includegraphics[width=0.47\textwidth]{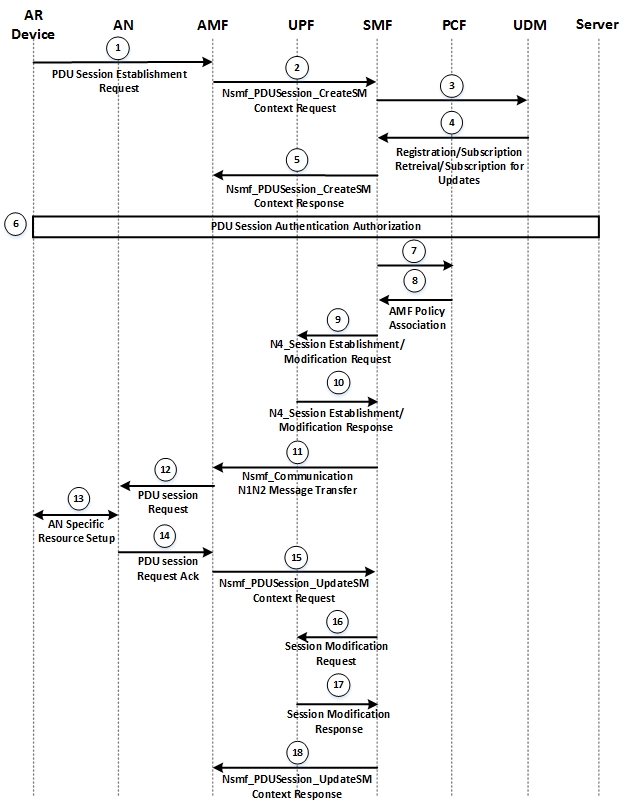}
  \caption{Message sequence chart for session establishment between AR device and server }
  \label{fig_session}
\end{figure}

After completing the registration process, AR device has to establish a data session with the image processing server to enable continuous data transfer. We define the Protocol Data Unit (PDU) session establishment procedure between AR device and the image processing server based on 3GPP specifications~\cite{procedures}. Figure \ref{fig_session} illustrates the message sequence required for PDU session establishment process.

Here, AR device initiates the process by sending PDU session establishment request to AMF via gNB. AMF then sends a request for a new session creation to SMF. In the next step, SMF registers with UDM, subsequently UDM stores data related to the session. After that SMF sends the Response to AMF. Then, PDU session authentication/authorization process occurs by exchanging messages between AR device, gNB, AMF, SMF, UPF and Server. Once this step is completed, SMF works with PCF for policy association for the session. Then UPF and SMF exchange the session establishment/modification request and the respective response. Message transfer from SMF to AMF allows AMF to know which access towards the AR device to use. AMF then sends the PDU session ID information to gNB so that gNB can work with AR device for the gNB specific resource setup. After that gNB sends the acknowledgement for the PDU session request to AMF. Based on that, AMF sends request regarding PDU session update to SMF and SMF then requests UPF for session modification. Once SMF received the response from UPF, SMF finally sends the response for PDU Session update to AMF completing the PDU session establishment process.

After successful completion of above steps, AR device can send a continuous data stream to the server and retrieve the augmentations sent by the server. Entities involved in this data transfer process are AR device, gNB, UPF and the server. Based on the above steps, 5G network functions needed to cater the AR use case can be identified to derive uO architecture.

Network slicing proposes a way to create logical networks on a common infrastructure to enable different types of communication services~\cite{alliance2016description, zhang2017network}. Assuming uO requires multiple network slices to cater a particular use case (eg. AR use case), uO has to create network slices before any actual communication happens over a selected slice. 3GPP introduces three network slice management functions for creating and managing network slices~\cite{slicing}.

\begin{itemize}
    \item \textbf{Communication Service Management Function (CSMF):} Responsible for translating communication service related requirement to network slice related requirements. 
    \item \textbf{Network Slice Management Function (NSMF):} Derive network slice subnet related requirements from network slice related requirements and, responsible for management and orchestration of Network Slice Instances (NSI).
    \item \textbf{Network Slice Subnet Management Function (NSSMF):} Responsible for management and orchestration of Network Slice Subnet Instances (NSSI).
\end{itemize}

\begin{figure}[ht]
\centering
    \includegraphics[width=0.47\textwidth]{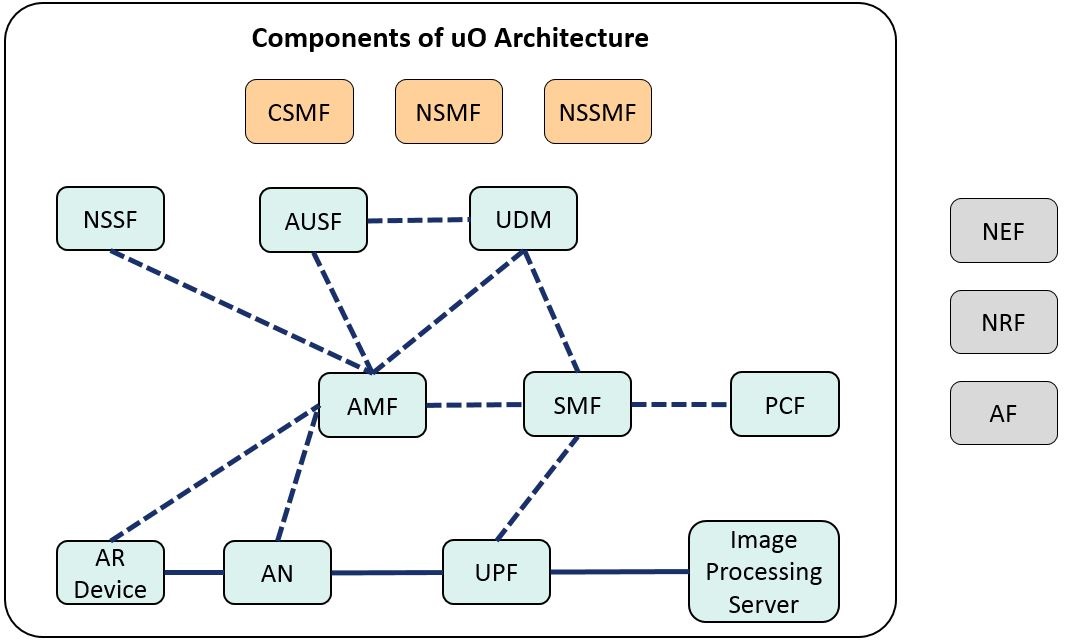}
  \caption{Proposed uO architecture for AR use case}
  \label{fig_arch2}
\end{figure}

For uO to create and manage multiple network slices, these three slice management functions should also be there in the architecture. When there are multiple slices available to facilitate any communication, the best fitting slice must be selected before the communication begins. This is done by Network Slice Selection Function (NSSF), which is an obligatory element in the uO architecture that supports multiple network slices. Figure \ref{fig_arch2} represents the proposed uO architecture for AR use case. For the sake of clarity, all the network functions in the original 5G architecture are illustrated in the Figure \ref{fig_arch2}, but NEF, NRF and AF are not necessary in this AR use case.

\section{Experimental Setup}
\label{sec:experimentalsetup}

\begin{figure}[ht]
\centering
    \includegraphics[width=0.47\textwidth]{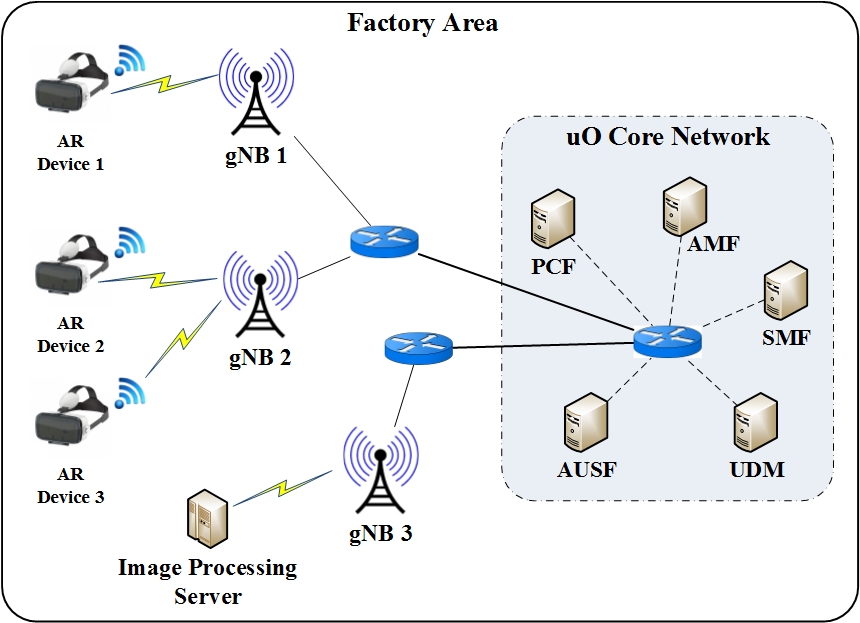}
  \caption{Deployment model for uO serving the factory }
  \label{fig_uo_sim}
\end{figure}

We consider two deployment models for our simulations for AR use case communications. In the first model, a factory is served by a local 5G network deployed by the uO. Factory owns the AR devices and the processing server. It is assumed that the server is located in a different cell site within the factory premises. Communication is facilitated by a 5G network deployed inside the factory premises. Core network of uO is also located inside the factory. This setup is depicted in Figure \ref{fig_uo_sim}.

\begin{figure}[ht]
\centering
    \includegraphics[width=0.47\textwidth]{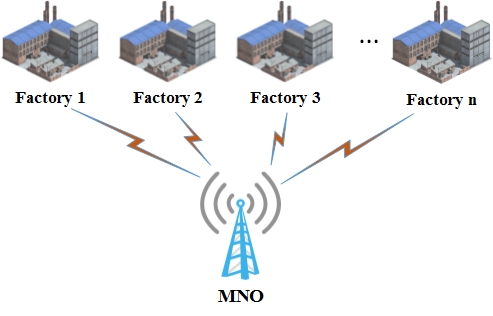}
  \caption{MNO's service for \(N\) factories having AR use case}
  \label{fig_fac}
\end{figure}

\begin{figure}[ht]
\centering
    \includegraphics[width=0.47\textwidth]{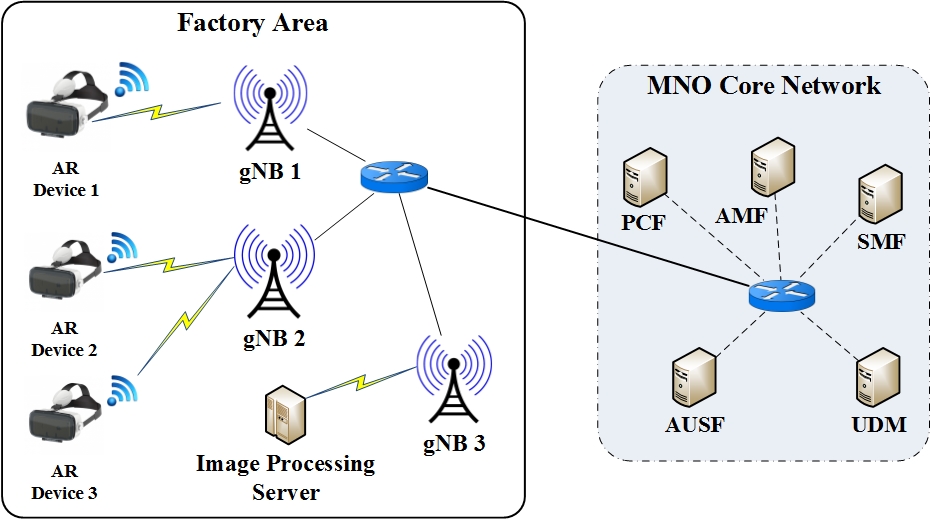}
  \caption{Deployment model for MNO serving the factory }
  \label{fig_mno_sim}
\end{figure}

In the second model, we assume that the entire factory is covered by a 5G network deployed by an MNO. AR devices and the processing server are owned by the factory. Server is located within the factory. We consider that the MNO is simultaneously serving total of \(N\) such factories having AR use cases. Each factory having similar network setup and  similar requirements as seen in Figure \ref{fig_fac}. Core network of MNO is located outside the factory. Figure \ref{fig_mno_sim} illustrates the MNO based model serving for the AR use case of a given factory.

Without loss of generality, we have assumed that uO's processing power is 1/\(N\) of the possessing power of an MNO. It validates the fact that MNO's resources are equally divided among \(N\) uOs in each factory.

For each uO and MNO based models, we simulate two of the above procedures namely the registration process and the data transfer between AR device and the image processing server. For the simulations, we use Omnet++ discrete event simulator~\cite{omnet} with INET framework installed~\cite{inet}. First step, the AR device registration process, has one AR device connected to a small cell base station, which then connects to 5G core. In this case, we model all the information flows from the AR device to Access Network~(AN), communication between the AN and the 5G core and back, and the information flow between the network functions which is needed for the registration process.

Data transfer process is the next step considered in our experiments. AR data stream travels through UPF of MNO/uO core via the 5G AN and then routes back to the image processing server inside the factory. The server then does the processing, generate the augmentations and initiate the data transfer to the AR device via UPF. This is modeled as a continuous data stream between the AR device and the image processing server.

Table \ref{table:gen_par} outlines the general simulation parameters whereas the variable parameters are explained at each experiment. Latency of AN is based on the 3GPP study on next generation access technologies~\cite{access} and we assume that both MNO and uO access networks serving the AR use case have similar properties. We take backhaul as a fiber connection and the latency parameters are selected based on a study of 5G backhaul challenges~\cite{jaber20165g}. Image processing server delay is based on 3GPP study on communication for automation~\cite{automation}. For each experiment, we measure the end-to-end~(E2E) latency of the communications.

\begin{table}[ht]
  \begin{center}
    \caption{ General simulation parameters}
    \label{table:gen_par}
    \begin{tabular}{|p{5.5cm}|c|} 
     \hline

      \textbf{Parameter} & \textbf{Value}\\

      \hline
      Latency between AR device and AN & 0.5 ms~\cite{access}\\
      Latency between AN and core network&~~0.05 ms per km~\cite{jaber20165g}\\
      Image processing server delay & 30 ms~\cite{automation}\\
      Distance to uO core network & 500 m\\
      Number of factories served by MNO (\(N\))  & 10\\
      \hline
    \end{tabular}
  \end{center}
\end{table}


\section{Results Analysis and Discussion}
\label{sec:discussion}

We run several experiments to study the performance of AR use case under uO and MNO network deployment models.

\subsection{AR Device Registration}\label{ARR}

\begin{figure}[ht]
\centering
    \includegraphics[width=0.47\textwidth]{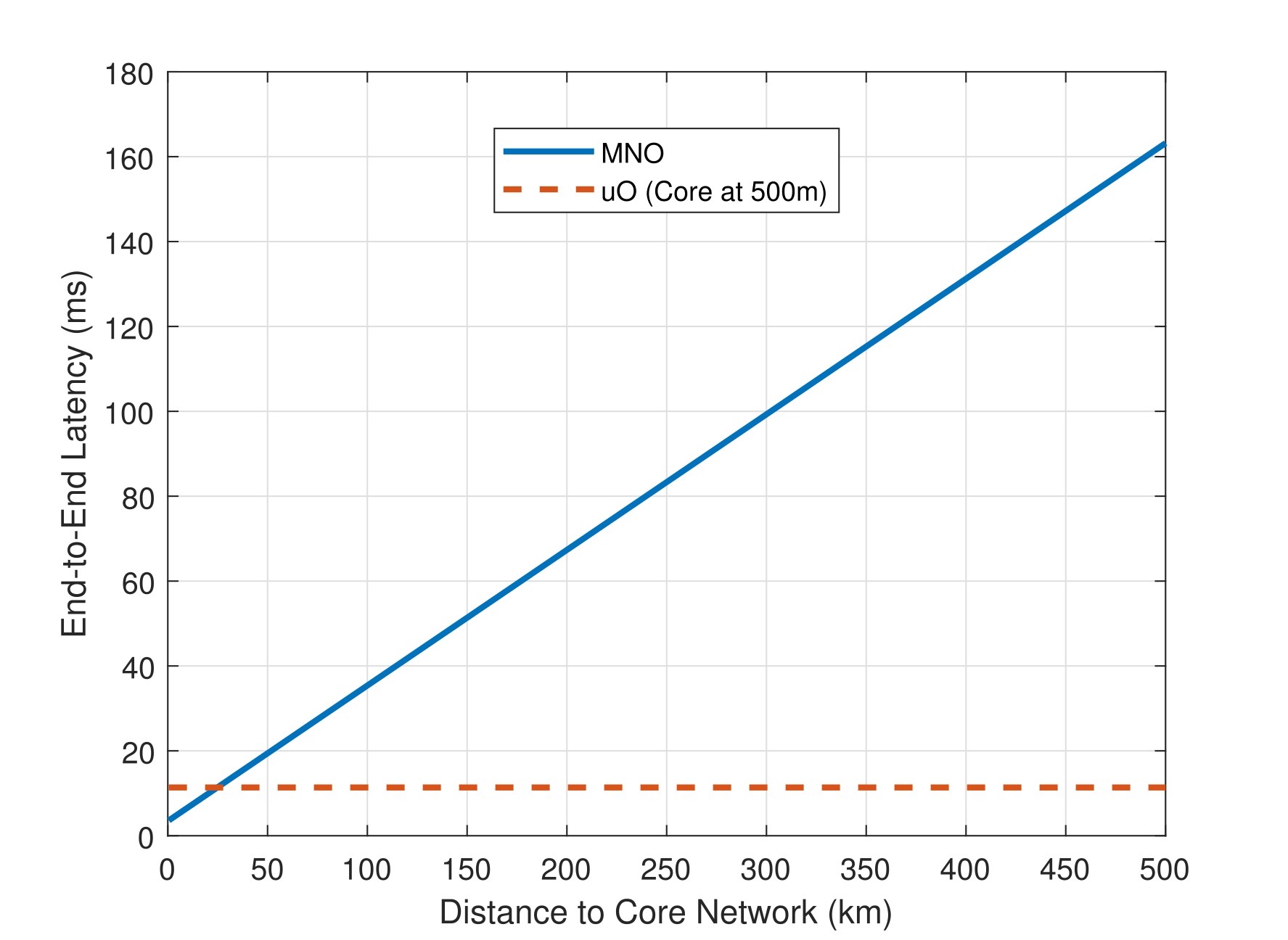}
  \caption{E2E latency of AR device registration process with respect to distance }
  \label{fig_lat_dis}
\end{figure}

We first simulate the registration process and observe the E2E latency with respect to distance to the core network of MNO. We use the parameters shown in Table \ref{table:gen_par} and vary the distance from 500 m to to 500 km in 50 km intervals. Results of the experiment are shown in Figure \ref{fig_lat_dis}. E2E latency of the 5G uO also illustrated in the Figure \ref{fig_lat_dis} assuming uO core network is located at 500 m. For uO setup, NF processing delay is taken as 1 ms, which is 10 times of the NF processing delay of MNO. 

E2E latency exhibits a linear increment with respect to distance to core network. Increment is approximately 0.32 ms per 1 km. Registration process has multiple round trips between AR device and the core network, causing the increase in latency. Even with 10 times higher NF processing time, 5G uO can still support a very low latency compared to an MNO having core network at a large distance. If the AR use case to be served by an MNO, then the core network should be in close proximity to factory premises to get low E2E latency as uO provides. It should be closer than 18.21 km as seen in Figure \ref{fig_lat_dis}. This is not a practical implementation because MNO is serving 10 factories and those factories usually are in diverse geographical areas.

\begin{figure}[ht]
\centering
    \includegraphics[width=0.47\textwidth]{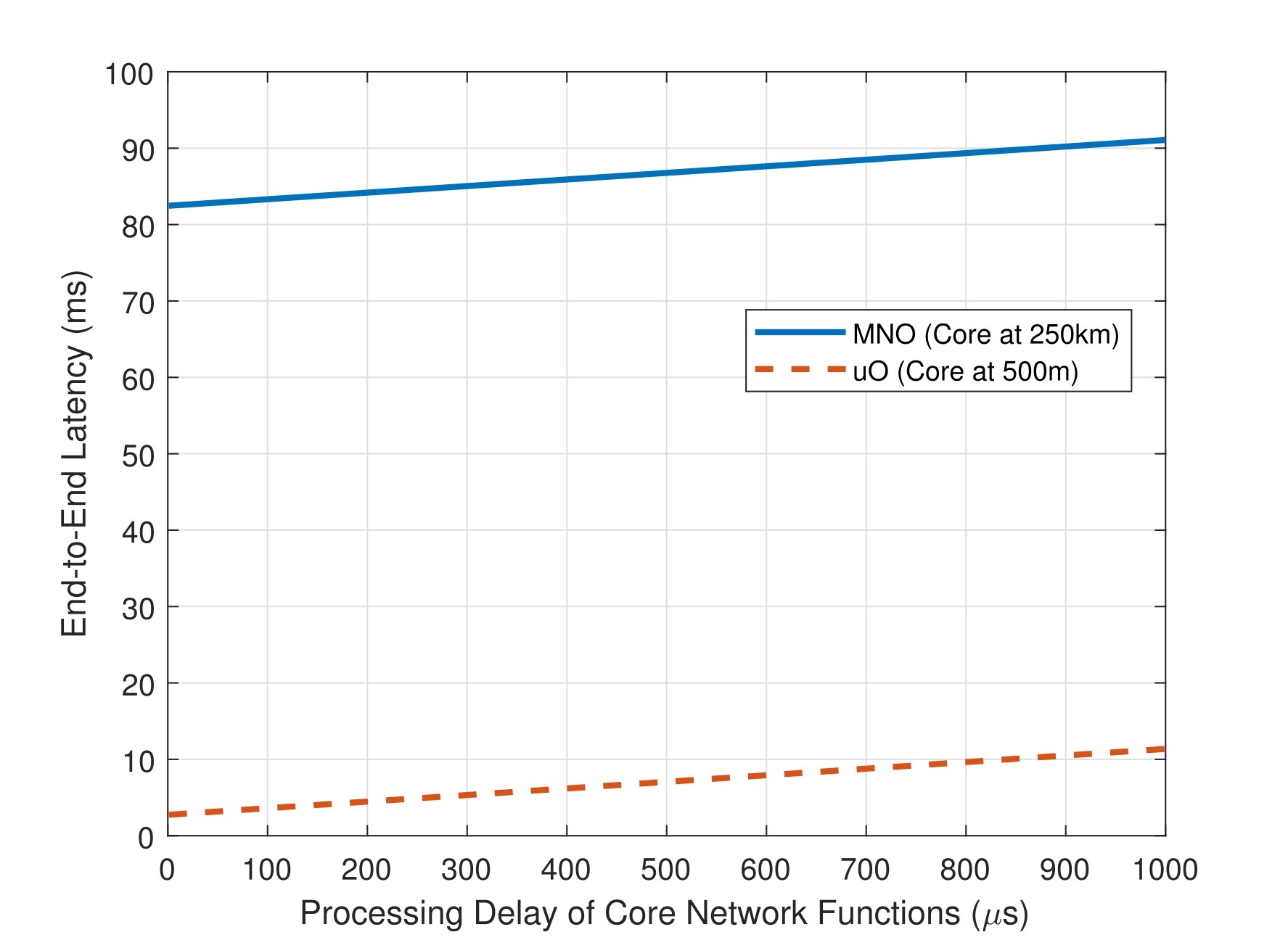}
  \caption{E2E latency of AR device registration process with respect to NF processing delay }
  \label{fig_lat_proc}
\end{figure}

Next, we observe the E2E latency with respect to NF processing delay. We keep the core network distance of uO at 500 m and MNO at 250 km. E2E latency obtained by varying NF processing delay from 1 $\mu$s to 1 ms is shown in Figure \ref{fig_lat_proc}.

E2E latency varies linearly with respect to NF processing delay for both uO and MNO. Latency increases approximately 8.63 $\mu$s when NF processing delay is increased by 1 $\mu$s. It is observed  that uO always performs better than MNO for a given NF processing delay. Reason for the behavior is that propagation delay due to core network distance is dominant than the NF processing delay.

In the actual implementation, uO 5G network is setup to serve only for AR use case because of the fact that uO is providing a tailored service to specific locations. However, MNO needs to handle 10 factories and different traffic forms such as mobile broadband traffic from users. Therefore, there is a high probability of having high NF processing delay in MNO case compared to uO case. 


As per the message transfer procedure of the registration process, E2E latency for the registration process can be expressed as follows,

\begin{equation}
    L_{reg} = k_{1}\:.T_{access} + k_{2}\:.T_{backhaul} + k_{3}\:.T_{N\!F}
\end{equation}

\noindent
where \( L_{reg}  \) is the E2E latency of registration process, \( T_{access}  \) is the delay from AR device to AN, \( T_{backhaul}  \) is the delay from AN to core network, \( T_{N\!F}  \) is the NF processing delay. \( k_{1}  \),  \( k_{2}  \) being the number of times where the registration message passes through access channel and backhaul respectively. \( k_{3}  \) is the number of times where the message is processed at a network function. Since \( T_{backhaul}  \) depends on the distance from AN to core network,

\begin{equation}
    T_{backhaul} = k_{4}\:.D_{backhaul}
\end{equation}

\noindent
where \( D_{backhaul}  \) is the distance from AN to core network and \( k_{4}  \) is a constant related to the communication over fiber channel. Therefore, for a given \( L_{reg}  \), distance to MNO core network can be calculated as,

\begin{equation}
    D_{backhaul} = \frac{L_{reg} - k_{1}\:.T_{access} - k_{3}\:.T_{N\!F} }{k_{2}\:.k_{4}}
\end{equation}

Moreover, NF processing delay varies based on two main factors, i.e. operator resources and network traffic load. NF processing delay is proportional to network load factor while inversely proportional to operator resources factor.

\begin{equation}
    N\!F\:processing\:delay \propto \frac{network\:load}{operator\:resources}
\end{equation}

Therefore, we consider following 4 cases of different resource levels at MNO and identify the minimum distance to core network to obtain same E2E latency as uO supports. 

\begin{itemize}
    \item Case 1 : MNO resources = uO resources
    \item Case 2 : MNO resources = 10 x uO resources
    \item Case 3 : MNO resources = 100 x uO resources
    \item Case 4 : MNO resources = 1000 x uO resources
\end{itemize}

Here also, we take MNO is serving 10 factories while uO is serving for a single factory.

Table \ref{table:dis_reg1} shows the maximum distance from the factory for above four cases, where the MNO core network should be placed to achieve similar performance as uO. In the first experiment, NF processing delay of uO is set to 1 ms. For MNO it varies for each case because of the availability of different resource levels. When the resource level of MNO and uO is equal, MNO cannot cater the E2E latency supported by uO because MNO is serving ten factories and the resources are shared among ten factories. When MNO resources is 10 times of uO resources, MNO can support the same E2E latency at the distance of 500 m because this case is the same as uO serving one factory. At higher resource levels, it is observed that the MNO can achieve the same E2E latency provided by uO, with the core network being located at a certain distance but not too far from the factory premises. This is due to backhaul delay, which is prominent than NF processing delay. When the resources of MNO is increased from 10x to 100x of uO, MNO can establish the core at 18.21 km. When the resources of MNO is increased from 100x to 1000x of uO, MNO can have a distance advantage from 18.21 km to 20.52 km, which is less than the distance advantage it gained by increasing resource level from 10x to 100x. This means that the distance advantage MNO gains, is diminishing even with the increase of core network resources.

\begin{table}[ht]
  \begin{center}
    \caption{Distance to MNO core when uO  \( T_{N\!F}  \) = 1 ms }
    \label{table:dis_reg1}
\begin{tabular}{ |c|c|c|c|c| } 
\hline


\textbf{Case} & \textbf{NF Proc. Delay of MNO} & \textbf{D\textsubscript{\textit{backhaul}}}\\
\hline
Case 1 & 10 ms & --\\ 
Case 2 & 1 ms & 500 m \\ 
Case 3 & 0.1 ms & 18.21 km\\
Case 4 & 0.01 ms & 20.52 km\\
\hline
\end{tabular}
\end{center}
\end{table}

We consider two more scenarios with less uO operator resources.  NF processing delay for uO is taken as 10 ms and 100 ms because \( T_{N\!F}  \) is inversely proportional to resource availability. Table \ref{table:dis_reg2} shows the minimum distance to MNO core from factory for \( T_{N\!F}  \) of uO = 10 ms and Table \ref{table:dis_reg3} shows the minimum distance to MNO core from factory for \( T_{N\!F}  \) of uO = 100 ms.

\begin{table}[ht]
  \begin{center}
    \caption{Distance to MNO core when uO  \( T_{N\!F}  \) = 10 ms }
    \label{table:dis_reg2}
\begin{tabular}{ |c|c|c|c|c| } 
\hline
\textbf{Case} & \textbf{NF Proc. Delay of MNO} & \textbf{D\textsubscript{\textit{backhaul}}}\\
\hline
Case 1 & 100 ms & --\\ 
Case 2 & 10 ms & 500 m \\ 
Case 3 & 1 ms & 231.92 km\\
Case 4 & 0.1 ms & 255.07 km\\
\hline
\end{tabular}
\end{center}
\end{table}

\begin{table}[ht]
  \begin{center}
    \caption{Distance to MNO core when uO  \( T_{N\!F}  \) = 100 ms }
    \label{table:dis_reg3}
\begin{tabular}{ |c|c|c|c|c| } 
\hline
\textbf{Case} & \textbf{NF Proc. Delay of MNO} & \textbf{D\textsubscript{\textit{backhaul}}}\\
\hline
Case 1 & 1000 ms & --\\ 
Case 2 & 100 ms & 500 m \\ 
Case 3 & 10 ms & 2314.78 km\\
Case 4 & 1 ms & 2546.21 km\\
\hline
\end{tabular}
\end{center}
\end{table}

When the uO performance is low, then the MNO has the ability to establish the core network at larger distances to provide the same latency as uO provides. However, this is unrealistic as uOs have the ability to provide a case specific service which means that the uO performance is tailored for that specific service. Moreover, processing delay in 5G networks will be within $\mu$s than ms range in order to support low latency applications.  Hence, we can consider uO \( T_{N\!F}  \) = 10 ms and uO \( T_{N\!F}  \) = 100 ms scenarios as non realistic ones.

\subsection{E2E Data Transfer}

\begin{figure}[ht]
\centering
    \includegraphics[width=0.47\textwidth]{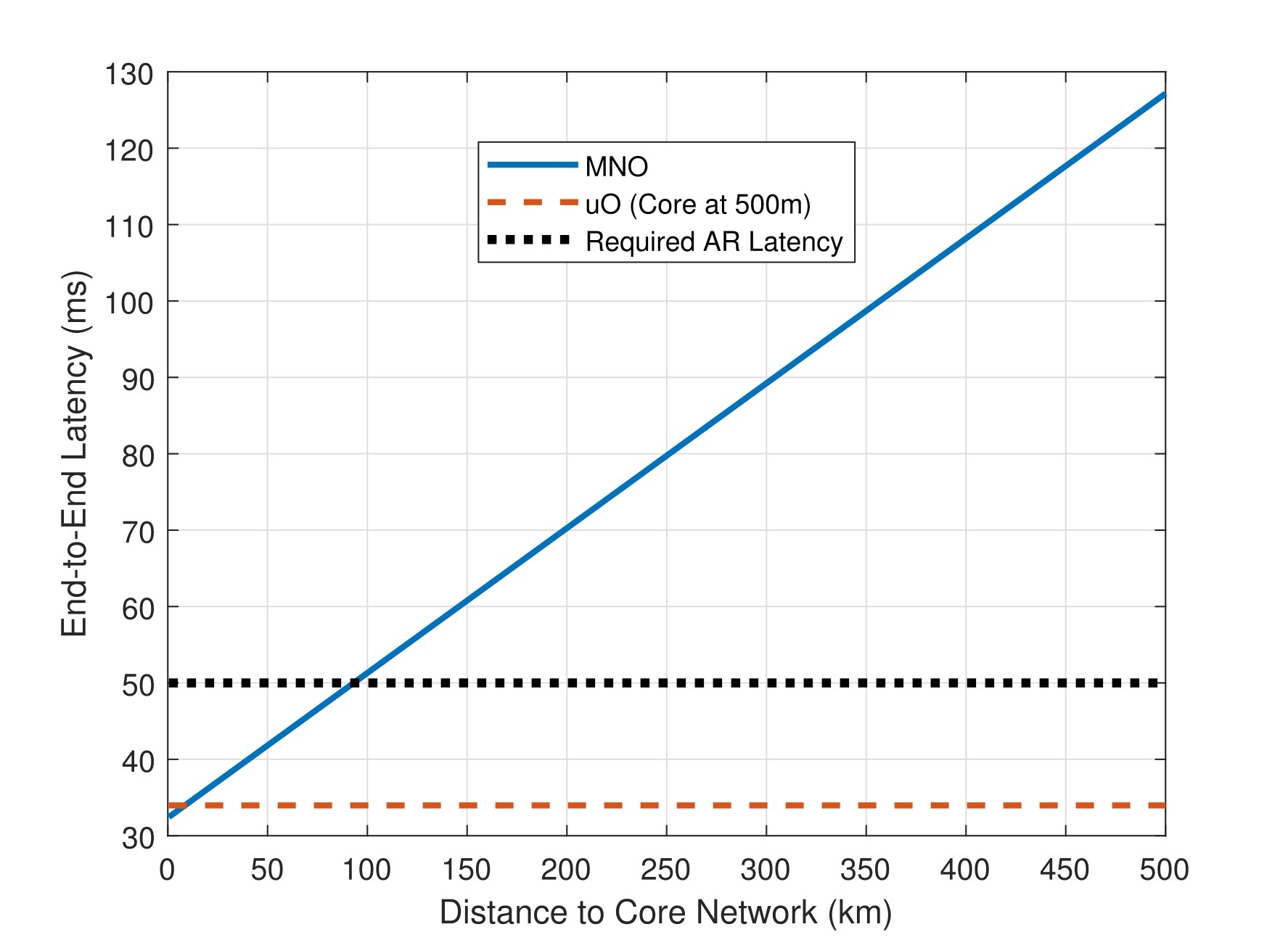}
  \caption{E2E latency of data transfer between AR device and server }
  \label{fig_lat_data}
\end{figure}

Here, we simulate the data transfer process from AR device to image processing server and back. According to ~\cite{automation}, this communication should be completed within 50 ms in order to avoid cyber-sickness. Therefore we try to identify the distance to core network to satisfy the latency requirement in an MNO served factory. We consider the NF processing delay of uO to be 1 ms and for MNO to be 0.1 ms. Results of the experiment are illustrated in Figure \ref{fig_lat_data}. Threshold latency of 50 ms and the uO performance with core network at 500 m are also depicted in the same figure for comparison.

Results show that if MNO is to serve the E2E latency requirement mentioned at ~\cite{automation}, its core network should be located approximately 92 km from the factory location. This requirement is difficult to satisfy when the ten factories are located in diverse geographical areas. Further, it shows that the uO can have far better performance than MNO, even though the NF processing delay of uO is 10 times higher than the MNO, making uO the favorable implementation option.

Similar to the registration process, we can express the E2E latency of data transfer process as,

\begin{equation}
    L_{dat} = k_{1}\:.T_{access} + k_{2}\:.T_{backhaul} + k_{3}\:.T_{N\!F} + T_{server}
\end{equation}

where \( L_{dat}  \) is the E2E latency of data transfer process, \( T_{server}  \) is the aggregated delay at the image processing server. For a given \( L_{dat}  \), we can calculate the distance to MNO core network as,

\begin{equation}
    D_{backhaul} = \frac{L_{dat} - k_{1}\:.T_{access} - k_{3}\:.T_{N\!F} - T_{server} }{k_{2}\:.k_{4}}
\end{equation}

As we explained in Section \ref{ARR}, we consider the four different resource levels at MNO and observe how far MNO can move its core network away from the factory premises, so that MNO can provide the same E2E latency as uO provides. Results of this experiment are outlined in Table \ref{table:dis_dat}. As in registration process, if both uO and MNO have same resources, MNO cannot cater the E2E latency supported by uO. When MNO resources is 10x higher than of uO, MNO provides similar E2E latency as uO with a core network distance of 500 m. With higher resource levels, MNO can get only a slight advantage for core network distance. MNO gets a 9.5 km advantage by increasing the resources from 10x to 100x, and a further 1 km advantage by increasing its resource level from 100x to 100x of uO, thereby making MNO a non-favorable choice.

\begin{table}[ht]
  \begin{center}
    \caption{Distance to MNO core when uO  \( T_{N\!F}  \) = 1 ms }
    \label{table:dis_dat}
    \begin{tabular}{ |c|c|c|c|c| } 
    \hline
    \textbf{Case} & \textbf{NF Proc. Delay of MNO} & \textbf{D\textsubscript{\textit{backhaul}}}\\
    \hline
    Case 1 & 10 ms & --\\ 
    Case 2 & 1 ms & 500 m \\ 
    Case 3 & 0.1 ms & 9.5 km\\
    Case 4 & 0.01 ms & 10.4 km\\
    \hline
    \end{tabular}
    \end{center}
\end{table}

As the second scenario we take a uO with \( T_{N\!F}  \) = 10 ms and obtain the E2E latency measurement for data transfer process and the result is 52.1 ms. Since the E2E latency required by AR use case is 50 ms, this scenario is not realistic. Therefore we do not further analyze core network location of MNO for this scenario.


\section{Conclusions}
\label{sec:conclusions}

The novel concept of micro-operator~(uO) enables a versatile set of stakeholders to operate 5G networks within their premises with a guaranteed quality and reliability to complement traditional Mobile Network Operator (MNO) offerings. In this paper, we analyzed the feasibility and performance advantage of using a uO instead of an MNO in a smart factory environment which supports industry 4.0 standards. 

We proposed a novel uO architecture based on 5G standards which is customized for factory environment. The architecture was discussed in terms of network functions and operational units which entail the core and the access networks in a smart factory environment. To realize the conceptual design, we conducted several experiments for an Augmented Reality~(AR) use case which will be heavily used in future factories.

The experiments revealed that a local 5G network established by the uO within the factory premises can cater low end-to-end latency for the AR use case compared to an MNO provided 5G network, where the core network is located outside the factory premises. End-to-end latency of the communication exhibits a significant increase over the distance between the core network and the factory. MNO can reduce the end-to-end latency by establishing the core network within close proximity (less than 92 km) to the factory or even inside the factory, however these deployments are not realistic because MNO usually provides services to multiple factories distributed geographically.

MNO can also have a different approach to reduce the end-to-end latency by increasing the computational resources at the core network, thereby effectively reducing the processing time within the core network functions. However, advantage MNO can yield by increasing resources is comparatively low because the dominant factor causing the end-to-end latency is the propagation delay due to distance to core network.

In a 5G uO  served factory, the AR data stream stays within the factory premises because the core network is inside a confined environment. This ensures more secure communication between AR devices and the server.

In future, we consider few more Industry 4.0 use cases such as mobility of robots and alarm management using sensor networks to study the impact of uO architecture.

\addtolength{\textheight}{-12cm}

\section*{Acknowledgement}
This work is supported by Business Finland in uO5G project and Academy of Finland in 6Genesis Flagship (grant no. 318927).


\bibliographystyle{IEEEtran}
\bibliography{references}

\vfill

\end{document}